# The Societal Impact of Big Data: A Research Roadmap for Europe

Martí Cuquet[a], Anna Fensel[a]

[a] *Semantic Technology Institute, Department of Informatics, University of Innsbruck, Technikerstraße 21a, 6020 Innsbruck, Austria*

**Abstract**

With its rapid growth and increasing adoption, big data is producing a substantial impact in society. Its usage is opening both opportunities such as new business models and economic gains and risks such as privacy violations and discrimination. Europe is in need of a comprehensive strategy to optimise the use of data for a societal benefit and increase the innovation and competitiveness of its productive activities. In this paper, we contribute to the definition of this strategy with a research roadmap to capture the economic, social and ethical, legal and political benefits associated with the use of big data in Europe. The present roadmap considers the positive and negative externalities associated with big data, maps research and innovation topics in the areas of data management, processing, analytics, protection, visualisation, as well as non-technical topics, to the externalities they can tackle, and provides a time frame to address these topics in order to deliver social impact, skills development and standardisation. Finally, it also identifies what sectors will be most benefited by each of the research efforts. The goal of the roadmap is to guide European research efforts to develop a socially responsible big data economy, and to allow stakeholders to identify and meet big data challenges and proceed with a shared understanding of the societal impact, positive and negative externalities and concrete problems worth investigating in future programmes.

**1 Introduction**

The volume of data is growing exponentially, and is expected to reach the tens of zettabytes in 2020, of which a third is expected to be valuable if analysed, and about 40% will require protection [1]. The acquisition, analysis, curation, storage and usage of such big data may result in effects experienced by third parties that had no direct involvement in the activity itself. These externalities—positive if the action causes a positive effect or benefit to the third party, negative if it causes cost or harm—arise from decisions, activities or products by stakeholders such as industry, researchers and policy-makers.

The present document contributes to the formulation of a strategy to define research and innovation efforts necessary for the realisation of a European big data economy by capturing and addressing the positive and negative societal externalities associated with the use of big data. It complements the technical challenges already identified [2] by taking a special focus on societal impacts, skills development and standardisation, and has been developed in parallel with a policy roadmap in the context of a multi-disciplinary study of the societal impacts of big data in seven European sectors aimed to define a roadmap and create a community that address and optimise these impacts [3].

The term big data has received numerous definitions [4,5]. To develop the roadmap, we considered as a working definition that big data is that which uses big volume, big velocity, big variety data assets to extract value (insight and knowledge), and furthermore ensures veracity (quality and credibility) of the original data and the acquired information, that demand cost-effective, novel forms of data and information processing for enhanced insight, decision making, and processes control. Moreover, those demands are supported by new data models and new infrastructure services and tools which are able to procure and process data from a variety of sources and deliver data in a variety of forms to several data and information consumers and devices [6].

The work presented here is the continuation of a series of case studies, analysis, expert focus groups and workshops conducted within the BYTE project funded by the European Commission. A total of seven sectors were considered as case studies of big data practices to gain understanding of the economic, legal, social, ethical and political externalities involved in them. They comprised crisis informatics [7], culture, energy [8], environment [9], healthcare, maritime transportation and smart city. A horizontal analysis of the societal externalities encountered in the case studies was conducted to identify how these externalities are connected to big data practices and to each other, to then evaluate and recommend how to address them (see, e.g. Ref. [10] for an analysis with a focus on data protection). Based on that, the vision statement for the BYTE project was presented [11]. During the preparation of the roadmap, two more workshops have been held to obtain feedback on the roadmap draft, validate its findings and further extend the results to a broader range of stakeholders.

The roadmap, together with the community being built around it, focuses on giving good practice messages about societal issues in big data. We consider the positive externalities, negative externalities and social impacts associated with big data, map research and innovation topics in the areas of data management, processing, analytics, protection, visualisation, as well as non-technical topics, to the externalities they can tackle, and provide a timeframe to address these topics and a prioritisation of them.

We have adopted a multilayered approach that accounts for what research will develop the necessary skills, contribute to standardisation and deliver social impact to capture positive externalities and diminish negative ones in the economic, social and ethical, legal, and political areas. We have also considered how such externalities affect different sectors, and which research topics are more relevant to each of these sectors. The originally planned time horizon for the present roadmap, 2020, has been extended to account for at least the upcoming 5 years after the roadmap presentation.

The present roadmap is expected to guide European policy and research efforts to develop a socially responsible big data economy. We also expect to contribute to the Big Data Value Association activities and priorities [12] by bringing a societal analysis of big data impacts, and to contribute to the creation of a multidisciplinary big data community around the BYTE results that includes as well NGOs, non-profit organisations, government (and especially local government) organisations, civil society organisations and citizens.

In the BYTE vision, three large-scale problems in European big data policies, aligned with the three identified trends, became apparent [11]:

1. European policy may be unprepared for the positive and negative impacts of a rapid technological transition towards big data. Among the considered case studies, the healthcare, shipping, crisis informatics and environment sectors may be ill-prepared for such a rapid transition, while the smart city, energy and culture are already living or anticipating it.

2. European policy may be poorly equipped for changes in the hegemony of big data, with the considered case studies split between being well-prepared for a future with few big players in big data (smart city and shipping) or anticipating to work with a diverse set of actors (healthcare, crisis informatics, environment, energy and culture).

3. European policy setting needs to be prepared to address both open, public data sources and closed, proprietary protections. Most of the sectors here considered are unprepared for an open regime, the only exception being the healthcare case study.

The big data transition entails preparing for a potentially major technological change. This preparation may require setting forth a legislative agenda, planning for appropriate research investment, and skills-force or capability planning for Europe as a whole. The transition may also have implication for European regions and their economic participation. The big data hegemony entails confronting the concentration of data in a few large players. This trend entails further issues of international economic competitiveness, issues of privacy, and issues of anti-trust legislation. The big data regime entails investigating who has control and access to data. This involves issues of transparency and intellectual property.

The rest of this paper is organised as follows. Section 2 presents the roadmap purpose and scope, and describes the methodology used to produce it. Section 3 identifies and discusses the current research requirements. First, the societal externalities that are to be addressed by the research topics are presented. The second subsection presents the five technical areas of research and innovation, each with several topics and recommendations. A sixth area discusses non-technical priorities. At the end of the section, the standardisation and skills requirements that are to deliver the expected societal benefits are presented. Section 4 presents the action plan that results from the requirements analysis. This includes a timeframe to address each research topic and how they contribute to industry sectors via societal impact, skills development and standardisation, and a discussion of best practices. Finally, Section 5 concludes the paper and presents final remarks.

**2 Scope and methodology**

*2.1 Roadmap purpose*

One of the primary goals of the BYTE project is to devise a research and policy roadmap that provides incremental steps necessary to achieve the BYTE vision and guidelines to assist industry and scientists to address externalities in order to improve innovation and competitiveness. The roadmap, together with the community being built around it, focuses on giving good practice messages about societal issues in big data, and in particular to the environment, healthcare and smart city sectors, which have been selected by the BYTE big data community as the ones to be addressed first.

The research roadmap focuses on what research, knowledge, technologies or skills are necessary in order to capture the economic and social benefits associated with the use of big data. It considers the positive externalities, negative externalities and social impacts associated with big data, maps research and innovation topics in the areas of data management, processing, analytics, protection, visualisation, as well as non-technical topics, to the externalities they can tackle, and provides a timeframe to address these topics and prioritisation of them. The research roadmap revolves around knowledge surrounding economic, legal, social, ethical and political issues, as well as standards, interoperability, development of meta-data, etc. It examines capacities and skills needed in computer science, statistics, social science and other industries or disciplines to enable European actors to take full advantage of the opportunities surrounding big data.

The present roadmap has been developed to be in alignment with the Big Data Value Strategic Research and Innovation Agenda (BDV SRIA) that defines the overall goals and technical and non-technical priorities for the European Public Private Partnership on Big Data Value [12]. We recommend to incorporate the results of the present roadmap into the BDV SRIA, in particular to expand the societal part and non-technical priorities, which are currently unbalanced with respect to the technical ones.

*2.2 Roadmap scope*

In this roadmap we present which types of research and innovation are needed to capture positive externalities associated with big data and diminish negative ones to obtain the best societal impact, develop the necessary skills and contribute to technology and data standardisation. The roadmap considers research and innovation in five technical areas (data management, data processing, data analytics, data protection and data visualisation) and presents which topics have the highest priority to impact the societal externalities in the upcoming 5 years. The roadmap is expected to guide European policy and research efforts to develop a socially responsible big data economy.

The roadmap has been developed following a multilayered approach [13] that accounts for skills development, standardisation and social impact of positive and negative externalities associated with big data, and links research and innovation topics to the targeted externalities and the sectors affected. Phaal, Farrukh and Probert [13] propose a T-Plan fast-start approach to technology roadmapping that is primarily developed for use from a company perspective, but can be customised for a multiorganisational use of a group of stakeholders, and it has been explicitly done so in the context of disruptive technological trends.

The time horizon for the present roadmap is 2021. We have also included a mid- and long-term timeframe to be addressed after 2021. We defined four top layers, sometimes labelled as *know-why* [13], that encapsulate the organisational purpose and correspond to the BYTE externalities that the roadmap is intended to impact and potentiate (if positive) or diminish (if negative). The externalities are arranged in four areas and 18 coarse-grained externalities. In addition to these purpose layers, we also considered how this research impacts the different industry sectors studied in the BYTE project and beyond. These layers represent the society pull in the roadmap. We further defined six bottom layers, also known as *know-how*, corresponding to the five-technical and the non-technical research and innovation areas, or resources, that are to be addressed to meet the demands of the top layers, and that encode the technology push. Finally, the middle layers of the roadmap connect the purpose with the resources to deliver benefits to stakeholders, i.e. represents the *know-what*. This includes the skills development, standardisation efforts and societal impact that the research and innovation actions contribute to.

The roadmap aims to deliver the vision for Europe. Nevertheless, to create it, we also took into consideration the Big Data research and innovation efforts and roadmaps from the whole world, and we are placing the roadmap in the international context, emphasizing the key visible similarities and differences to the other countries of the world [14]. We relate to the three aspects of the roadmap: its construction process, the addressed research and innovation topics, as well as the prioritised public and private sector areas.

For the present roadmap construction process, similar instruments to the ones used elsewhere have been applied, like workshops, interviews, stakeholder consultations. For example, details on the inputs for the US roadmap construction process are provided in its description [15]. As with the US and other roadmaps, national stakeholders have been consulted, in order to focus on the topics, which are most crucial for Europe, and where the European research and development is most competitive.

Regarding the selected research and innovation topics, Data Analytics direction is essentially present ubiquitously in the world's Big Data roadmaps. Especially, it tops the lists for the countries that have access to large amounts of data, such as for example the US, that have large quantities of the users' and companies' data from the Web, and Asian countries, that generally have an access to massive amounts of data, originating from the Internet of Things. Machine Learning and Deep Learning are also prominently present in a number of roadmaps, e.g. of USA and China [15,16]: these research fields are related to data analytics and are relevant for applications such as recommendation and prediction. Open Data, its availability and role in making the public sector more transparent and efficient, have also substantially spread increased in the last years: it has been largely driven by the US and followed by developed countries on all continents. As large data brings large responsibility and eventually has an effect on individual lives of people, privacy and placing the users in control of their own data is mentioned throughout the roadmaps and acted on explicitly in the legislation base of the many of countries: USA [15], Russia, Japan, etc. Privacy-aware access to Big Data also has been identified as a high priority direction in our roadmap.

Finally, many world's roadmaps, including ours, list specific industries and sectors where the developments are most crucial and expected for the country for which the roadmap has been produced. While in our roadmap the sectors of health, environment and smart city came in the first priority, other countries' priorities only are partly overlapping. For example, the USA is explicitly supporting health care, education, homeland security, law enforcement and privacy law in public sector, as well as supporting the consumers and enterprises, advertising-supported industry, and data services in the private sector [15], while the most recent US Big Data case studies include access to credit, employment, higher education, and criminal justice [17]. And in China, for example, the manufacturing industry, as well as the environment and decrease of pollution, productivity of public sector and optimisation of transport are appearing in the high priorities [14,16].

*2.3 Roadmapping process and methodology*

The roadmapping process built upon previous work within the BYTE project, and hence slightly deviated from common roadmapping method proposals and guidelines, which usually incorporate the creation of a vision and suggest conducting interviews with experts and holding several focus group discussions and workshops with relevant stakeholders. Such parts of the roadmapping process are already embedded in the BYTE project work and timeline and produced a number of deliverables that report the project findings and outcomes. The most relevant deliverables for the present roadmap are the horizontal analysis [18], the evaluation of externalities [19], and the vision documents [11,20]. Thus, standard roadmapping processes were adapted to review such deliverables and incorporate their findings into the relevant parts of the roadmapping, rather than redoing the tasks.

The development process of the present research roadmap was done in three phases, as shown in Figure 1. In the first phase, a purpose and scope statement was developed to guide and maintain the focus throughout the roadmap development process. This phase also included baseline research to identify stakeholders and relevant sectors beyond those studied by the BYTE project. In the second phase, the vision in [11] was summarised and clearly restated with a special focus in the topics of the research roadmap. This vision was subsequently amended to incorporate the project reviewers' recommendations and community feedback. In the third phase, we mapped how the research and innovation topics identified in the first phase may be used to address societal externalities, and analysed how they may impact society and contribute to standardisation and skills development in order to capture the positive externalities. The relevance of each externality to the BYTE sectors and the mapping of research topics to externalities and sectors was assessed by a review of the case study reports [21] and complemented with an analysis of big data initiatives and external studies [2,12,22–25] to include each significant contribution to the roadmap.

To prioritise and bring the themes together on a time-basis [13], we conducted the *BYTE Big data research roadmapping workshop* on 1 July 2016, collocated with the European Data Forum 2016 that took place in Eindhoven, the Netherlands, with a total of 26 senior level participants from academia (11), SMEs (8), large companies (3), public organisations (3) and certification bodies (1), and 11 European countries (Austria, Belgium, Germany, Hungary, Ireland, Italy, the Netherlands, Norway, Spain, Sweden, and the United Kingdom). Participants were divided in 6 working groups to validate and further refine the priorities and research challenges, map them with their impact on societal externalities to complement the BYTE results and to temporally align and prioritise them. We incorporated and further expanded the action plan developed in the workshop at the end of the roadmapping process. The second draft was finally reviewed by the project members and advisory board, and validated and adjusted in the BYTE big data community workshop.

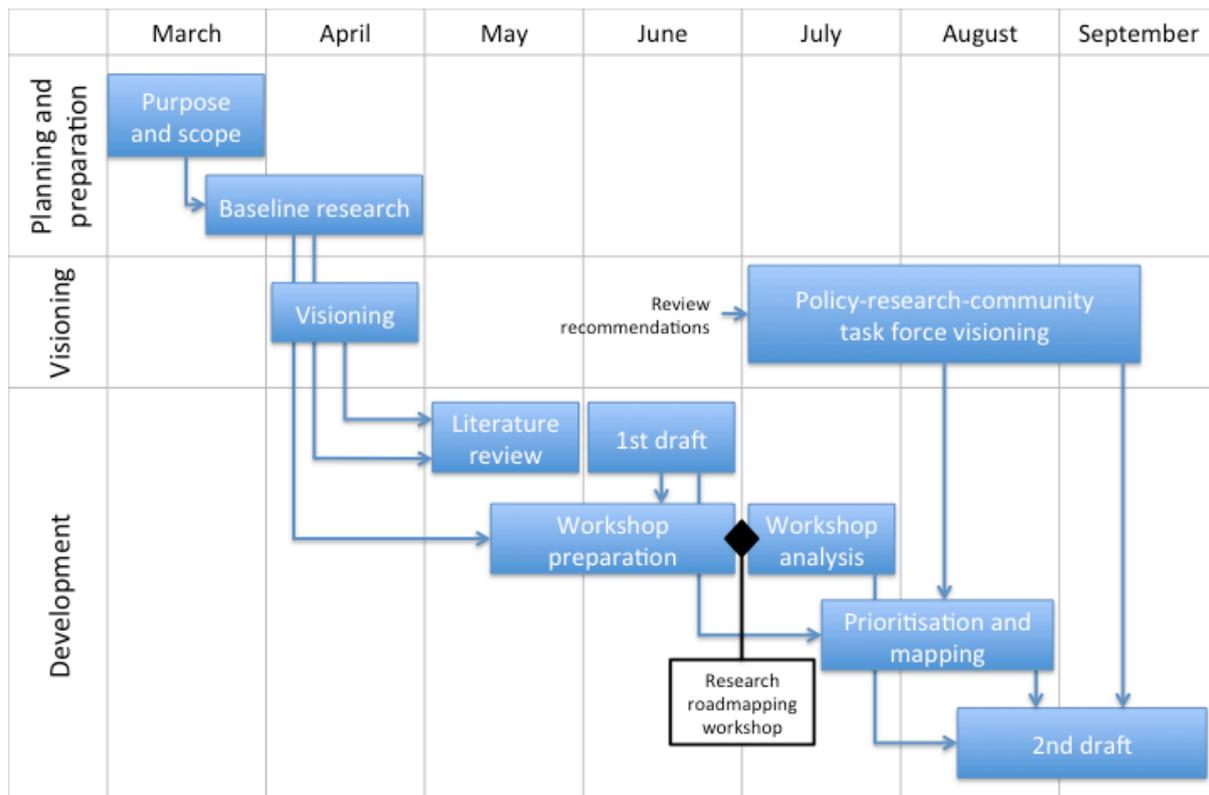

Figure 1. Research roadmap phases and timeline.

## 3. Requirements

This section presents the societal externalities and their relevance in several industry sectors, and the five technical areas of research and innovation are presented that can address them. A sixth area further discusses non-technical priorities. At the end of the section, the requirements in standardisation and skills development are also presented.

*3.1 Externalities*

The BYTE project identified and considered 73 societal externalities classified by the pairs of stakeholders involved (public sector, private sector and citizens) and their main topic (business models, data sources and open data, policies and legal issues, social and ethical issues, and technologies and infrastructures). For a thorough description and analysis of the externalities, we refer the reader to [14,21]. Throughout the project, we have used the following definition for externality [26]:

**Positive externalities** occur when a product, activity or decision by an actor causes positive effects or benefits realized by a third party resulting from a transaction in which they had no direct involvement.

**Negative externalities** occur when a product, activity or decision by an actor causes costs (or harm) that is not entirely born by that actor but that affects a third party, e.g. society. It is generally viewed as a failure of the market because the level of consumption or production of the product is higher than what the society requires.

As the boundary between internal and external is often arbitrary, we have in some cases extended the definition to include also the internal impact of a product, activity or decision.

The 73 externalities were simplified to 18, and grouped in four main areas [18]:

**Economic externalities:** improved efficiency, innovation, changing business models, employment, and dependency on public funding.

**Social and ethical externalities:** improved efficiency and innovation, improved awareness and decision-making, participation, equality, discrimination, and trust.

**Legal externalities:** data protection and privacy, intellectual property rights, and liability and accountability.

**Political externalities:** private vs. public and non-profit sector, losing control to actors abroad, improved decision-making and participation, and political abuse and surveillance.

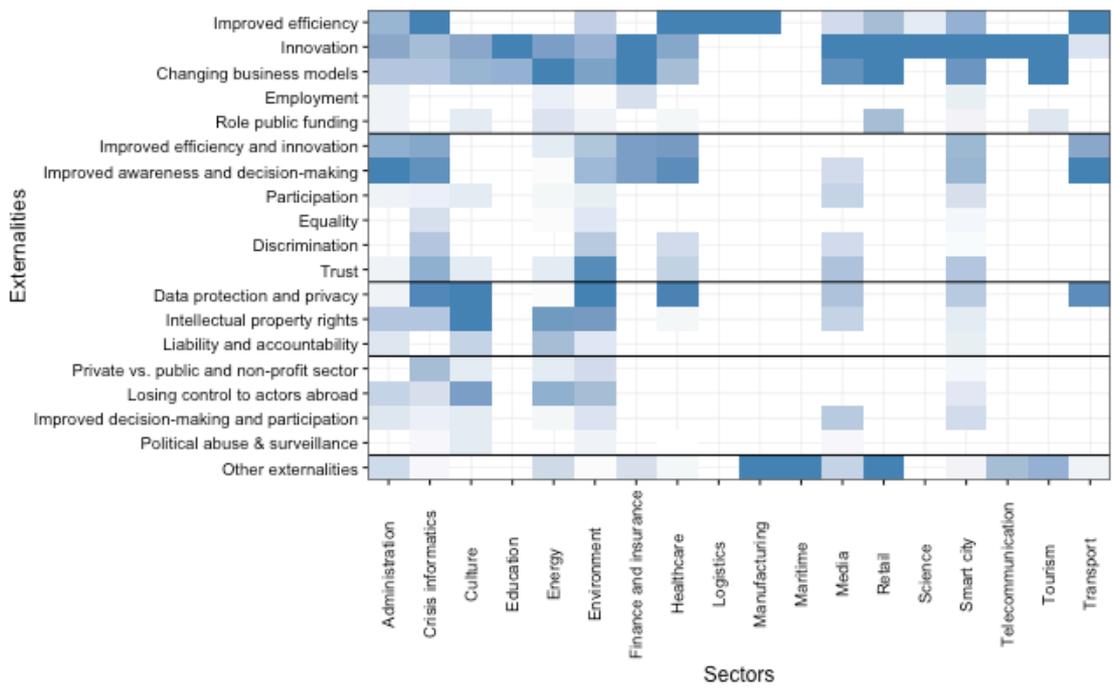

Figure 2. Relevance of externalities in different sectors, derived from BYTE analysis and the literature review. Relevance has been normalised by sector: darkest blue corresponds to the most relevant externality in the sector. Externalities are grouped in four areas (top to bottom): economic, social and ethical, legal, and political.

The relevance of each externality to the BYTE sectors has been assessed by the review as mentioned in Section 2.3, and normalised by sector. This relevance is summarised in Figure 2. It includes the sectors of crisis informatics, culture, energy, environment, healthcare and smart city from the BYTE case studies (see e.g. Ref. [3] for an overview of the cases and their associated externalities) and has been extended to extra sectors: administration, education, finance and insurance, logistics, manufacturing, maritime, media, retail, science, telecommunication, tourism and transport.

*3.2 Research and innovation topics*

In the development of the research roadmap, we have taken the approach outlined in the previous section and have considered the research and innovation topics of the BDV SRIA [12]. These topics have been further extended with observations and recommendations from the BYTE case studies, analysis and workshops and the contribution from community stakeholders, and aligned (in sections below) with the impact they have in society (i.e. how can they amplify the positive externalities and diminish the negative ones). The aim is that the roadmap could later be incorporated into the BDV SRIA, in particular to expand the societal part and non-technical priorities, which are currently unbalanced with respect to the technical ones.

In the following subsections we present the five areas of technical priorities and the non-technical priorities currently being considered at the Big Data Value Association (BDVA). To avoid overextending, the topics are not comprehensively described and we refer the reader to the BDV SRIA document [12]. Instead, in the following subsections we add further observations and subtopics to be considered in each of them that arose in the workshops and literature review, especially in regards to their societal impact.

3.2.1 Data management

**Handling unstructured and semi-structured data.** It was pointed out that multilingualism is still a challenge, especially in the processing of natural languages different from English. It has also been identified as an opportunity for Europe, as it already has the necessary skillset to address these requirements. There is also a need to develop easy-to-use reporting tools including semantic annotations that do not add extra work, e.g. to healthcare professionals [27]. This requirement is relevant in conjunction with the topic below.

**Semantic interoperability.** A relevant subtopic to be added is data integration and fusion. There are issues with format conversion that leads to intelligence losses. Currently, reengineering is the only way to recover intelligence, so there it is thus required to define new policies about how original files are kept, as well as to develop technology to ensure interoperability among different formats. In any case, linked data technologies need to be simplified in order to make them easily adoptable [28]. Also, semantic search, schema matching and mapping, and ontology alignment have to be addressed [29].

**Measuring and assuring data quality.** This topic should include transparency of the data collection process, and also meta information on the context and purpose of such collection. Approaches to compute the uncertainty of the results of algorithms need to be developed [29], in order to include evidence-based measurement models of uncertainty over data. Also, algorithms to validate and annotate data need to be developed [29]. Finally, funding agencies should require an explicit estimate of data curation and publication costs of high quality data [29].

**Data lifecycle.** Access to data is still put forward as one of the main challenges. It was mentioned that focus should be given on data that already exists rather than on data that needs to be created in the future, and thus data creation, with a focus on surfacing already existing data, is a priority within this topic. There is more data out there than what people realise, and it should be made easier to find [28]. This could be assisted by developing search engines for datasets with ranking, in order to drive owners to publish better datasets, following the improvement of websites that want to appear high in Google rankings [28]. Adaptive data detection and acquisition is needed in e.g. the finance and insurance sector [27]. Other relevant subtopics are data discovery, datasets crawlers, metadata, dataset ranking [28]. Data curation by demonstration, in analogy to programming by example or by demonstration, would also for the distribution and scalability of the system. Also, to be added here is the preservation and archiving of data. It has also arisen from several stakeholders that currently the biggest challenge is the variety of data. Within this topic, data citation, curation and preservation have been identified as additional relevant subtopics. In particular, standards for data citation are currently demanded. Understanding how data expires, what happens with historical data and how it is archived is important. In relation to this, the synchronisation of data and how to update extracted knowledge bases if the sources are changing should also be addressed [27]. Open data is also central to research itself. The Research Data Association is actively addressing how to build research data services with open linked data, for example in the publishing, referencing, citing and searching areas [30,31]. There are promising semantic languages and technologies that can be applied to such research data services (e.g. linked services, linked data, Schema.org), which can as well assist in multichannel research dissemination [32,33].

**Data provenance, control and IPR.** It has been highlighted that certain kinds of data attract new rights and require new rights statement initiatives. This has also a further impact in its implications within linked open data, and is particularly relevant for media data, where the digitalisation of an object (other than text) was put forward as an example. Data licensing and ownership techniques are still under development, and especially crucial for Internet of things applications, where data is distributed among different physical locations and where often the appliance and software manufacturers are the organisations that grab the data. Data curation depends on mechanisms to assign permissions and digital rights at the data level and to provide context through data provenance [29]. New theoretical models and methodologies for data transportability under different contexts should be developed [29]. Decisions taken during the data curation process need to be captured, and models and tools to grant fine-grained permission management developed [29]. In this same direction one finds sandboxing and virtualisation techniques [34]. Nanopublications may impact the development of distributed science via semi-structured data and scientific statements [35]. Another open challenge is how to guarantee the integrity and confidentiality of provenance data [34].

**Data-as-a-service model and paradigm.** Relevant subtopics are licensing, ownership and marketplace. Research has to be devoted also to the extraction of value from data, particularly in terms of what data needs to be created for maximum value extraction. In this regard, also the estimation of data value both in the present and future is a relevant topic. More services that take advantage of open data need to be supported [19].

Also, within the area of data management, and in relation to **open data practices**, it has been pointed out that there exist several open data issues for registered companies and lack of harmonization across Europe. This is not restricted only to the legal and policy framework, but can be addressed also from a technical perspective. In general, industry agrees with the need to open data but finds difficulty to make it open (especially in orphan works), and asks for financial support to open data. We thus recommend to add an **open data and data creation** priority within the area of data management aimed at surfacing already existing data. Public funding should keep prioritising processes that support open data initiatives [28]. Tracking and recognition of data and infrastructure should be improved in academic research [29]. Options for this include recognising open dataset publication analogously to paper publication in journals.

### 3.2.2 Data processing

**Techniques and tools for processing real-time heterogeneous data.** This is particularly needed in the development of new tools for sensor data processing, especially in the manufacturing, retail and transport sectors [27], as well as in the energy sector. Social media mining is also relevant [27].

**Scalable algorithms and techniques for real-time analytics** such as stream data mining in contexts of a high volume of stream data coming from e.g. sensor networks or large numbers of online users [28], as well as real-time analysis of public transportation data [19].

**Decentralised and distributed architectures.** This includes efficient and scalable cryptographic mechanisms for the cloud (e.g. directory-based encryption, container-based encryption, manual encryption) and attribute-based encryption [34,36]. Distributed architectures should also be explored as an opportunity to keep sensitive data on user-governed devices.

**Efficient mechanisms for storage and processing.** To automate complex tasks and make them scalable, hybrid human-algorithmic data curation approaches have to be further developed [29]. Crowdsourcing also plays an important role. Energy-efficient data storage methods are also a crucial research priority [34].

### 3.2.3 Data analytics

**Improved models and simulations.** There is a need of better integration between algorithmic and human computation approaches [29]. Catchment techniques, recommendations and customer tendency research are also relevant for the retail sector [19,28], and simulations for resource allocation for the crisis informatics and smart city sectors [19]. In general, most models would extremely benefit by methods to correct sample bias [37]. This affects also the data collection and quality assessment processes.

**Semantic analysis.** Examples are sentiment analysis, a relevant subtopic when using social media data for the manufacturing or retail sectors [27], and entity recognition and linking [29].

**Event and pattern discovery.** To be added within this innovation topic, but also relate to the predictive and prescriptive analytics below, is the need to further investigate and differentiate between correlation and causation. In this direction, an evidence-driven, bottom-up approach has been put forward to first deduce correlations from evidence (e.g. using data from economic phenomena) and then develop means to estimate their correctness and completeness, such as the probabilistic likelihood that correlations are causal within error bounds [38]. Anomaly detection can be applied e.g. to detect deviations from traffic in a smart city [28]. Clustering of social media post can also be used to detect and gather real-time information in emergencies [28], especially in real time [37]. Another topic is pattern recognition on imaging device results [19].

**Multimedia (unstructured) data mining.** Sentiment analysis beyond the analysis of textual information needs to be addressed [19]. In this regard, it was raised during the workshop that there is a lack of tools to deal with multilingual sentiment analysis, and Europe is probably in the best position to tackle this challenge.

**Machine learning techniques, especially deep learning for business intelligence, predictive and prescriptive analytics.** This topic is already coupled in the BDV SRIA document with the priorities on visualisation and end-user usability. But even more importantly that such usability of the analytics results by non-data scientists, it has been recognised an urgent need of validated methodologies and standards behind the analytics on whose results decisions are to be taken, and that are easily identifiable and understandable by decision-makers. Also relevant for decision-making is to correctly assess the representation of data and possible data biases, as it may lead to biased decisions. An emerging trend is to use new sensor data for predictive analysis, e.g. in Industry 4.0 [37].

### 3.2.4 Data protection

**Complete data protection framework.** It is important to create "resources for using commoditized and privacy preserving Big Data analytical services within SME's" [24]. The major security challenges are now in non-relational data stores [34]. Also, granular access controls have to be developed that allow to share data on a fine-grained level [29]. New legal means have to be developed too to handle access to data and the permitted use of data that ensure that data protection is not an obstacle for big data practices [19]. This includes a better scalable transaction model in data protection law [19].

**Privacy-preserving mining algorithms.** Although further research is still needed in this area, there exist already interesting approaches that are however not well-known in industry. There is a need thus to disseminate these results and bring them to practice [24]. Special emphasis has to be done in the mining algorithms of social media [19].

**Robust anonymisation algorithms.** This includes the development of novel algorithms such as k-anonymity [39].

**Protection against reversibility.** Considerable research is required to better understand how data can be misused, how it needs to be protected and integrated in big data storage solutions [34].

3.2.5 Data visualisation

**End user centric visualisation and analytics.** Natural language interfaces, and interactive and easy-to-use data access and transformation methods need to be further developed and brought to commercial applications [29].

**Dynamic clustering of information.** This requires efforts for new and better data summarisation and visualisation, and user interfaces for parallel exploration [37] such as subjunctive interfaces [40].

**New visualisation for geospatial data.** Geospatial data can benefit too from the user interfaces for parallel exploration mentioned above [37].

3.2.6 Non-technical priorities

**Establish and increase trust.** Open government data is widely recognised as a method to increase trust and transparency [28], although this requires parallel actions to reduce the digital divide [19]. A solution is an increase of data journalists who are able to process and present such data to a wider audience.

**Privacy-by-design.** Transparency for users is still an issue, so privacy-by-design and similar by-design approaches are vital [28]. By-design approaches are generally seen as a solution to allow business to evaluate and analyse data, and in particular sensitive data, without needing too restrictive provisions to avoid profiling [24]. An example is the fine-grained control of digital rights [41]. As anonymising and de-identifying data might be usually insufficient in view of the amount of data that can be used for re-identification, the transparent handling of data and algorithms and company audits should be considered [34].

**Ethical issues.** It has also been pointed out that further discussion is needed regarding whether research that analyses human data should fall within the regulations of research based on human subjects. This is in line with the discussions presented by the Council of Big Data, Ethics and Society [23].

There is a demand from the scientific community to access data owned by companies for research purposes. Standards should be set to enable such sharing of data across sectors in a way that allows companies to contribute anonymised data to the scientific community without the possibility of backfiring, as has happened in past experiences [23]. Research is needed to quantify the risks posed by data science practices that rely on big data. This includes dealing with minimal individual risks that however affect a very large population and with privacy risks that depend on a highly varying privacy expectation of subjects in the same study [23]. Research is also needed to account for and mitigate the risks of sharing datasets that can be later combined with auxiliary datasets, thus e.g. increasing the risk of de-anonymisation. Research has already started in this direction [42].

Usage of publicly available, although illicitly obtained data sets is also a matter of controversy within the scientific community [23]. There is a need to establish at least best practices on how to approach this challenge. In industry, ethic processes and ethic review structures that work have to be developed and tested [23].

Bias is also a relevant ethical issue that needs further research. It is commonly implicit in big data processes that all data will eventually be sampled, although this is hardly ever true and there can indeed be a sample bias introduced by technical, economic or social factors [37]. Subjective bias can also be introduced in the data through the labelling of the data [28].

**Develop new business models.** Open source big data analytics have been proposed as a way to ensure that benefits remain in the EU [24]. However, moving beyond the Open Data Initiative to an interoperable data scheme to process data from heterogeneous sources is also seen as a way to foster and develop new business models [24]. Other novel models are pre-competitive partnerships where organisations that are typically competitors cooperate in R&D projects of certain data value chain steps, such as data curation, that do not affect their competitive advantage and public-private partnerships [29].

**Citizen research.** Crowdsourcing may be used to increase data accuracy [28] and scale data curation [29], among other applications. New methods are needed to route tasks to crowdsourcing participants based on their expertise, demographic profiles, and long-term teams, and develop open platforms for voluntary work [29]. Research is needed to better understand the social engagement mechanisms, e.g. in projects such as Wikipedia, GalaxyZoo [43] or FoldIt [44], which would amplify community engagement [29].

**Discrimination discovery and prevention.** Within this topic, or as a priority of its own, a more research on legal informatics and algorithm accountability is needed. This is especially relevant for IPR-related externalities.

*3.3 Skills development and standardisation*

In addition to the societal impact described by the externalities presented in Section 3.1, in this Section we discuss priorities that fall into skills development and standardisation efforts.

The need for educated people equipped with the right data skills has been extensively identified (see e.g. Refs. [2,24,45–48]. For example, the McKinsey report classifies the required skills in *deep analytical talents* to analyse the data, *data-savvy managers and analysts* to effectively consume the data and *supporting technology personnel* [45]. Similarly, the BDVA has also identified three profiles that partially overlap with the ones previously described: *data scientists*, *data-intensive business experts* and *data-intensive engineers* [12]. The European Data Science Academy project is addressing this challenge and has recently released a report that evaluates the skills gap and how to close it [49].

The BYTE case studies and analysis also identified and confirmed this need and recommend promoting big data in education policies [19]. This has been further confirmed in the big data research roadmapping workshop. It was noted that an interface between policy makers, society and industry is needed. This requires data-savvy professionals in all areas that have to take data-driven decisions, and not only the ubiquitously stated need of data scientists. Moreover, an emphasis has been put to integrate ethics education into the data science curricula. This is supported by similar recommendations elsewhere [23]. In the research area, priority should be given to simplify already mature technologies [28] and make them accessible to innovative businesses. In successive workshops, it has repeatedly been mentioned that the increase in data skills in the general public and in key expert positions could mitigate the large need of data scientists and engineers. Data-intensive policy makers are an example of a skill that was identified to be of high priority: more than the ability to deal with data, that of being able to correctly understand and interpret the models used to predict and make recommendation, and the type of data in which they are based. This includes more research into correlation vs. causation tools and the need of validated methodologies.

Another relevant aspect that came up is the digital divide and how open data, that supposedly benefits citizens in general, is actually more likely to increase the digital divide and produce social inequality, as data is effectively data "is only open to a small elite of technical specialists who know how to interpret and use it", and to those who can employ them [19]. This divide also affects the gender category: female leaders in industry and research are only a small percentage (e.g., 11% in ICT in Austria, compared to an already low 25% of average in other sectors [48]). An important action to decrease this divide is to promote data journalism to process, digest, and present the newly available open data to society.

Regarding standardisation, in general two types have been identified in alignment with the BDV SRIA: *technology* and *data standardisation*. It was though pointed out that an excess of standards, especially for interoperability, is not always useful and can lead to potentially negative changes in society, especially when they slow down innovation. In this case they should be replaced by best practice recommendations. Standardisation is in any case urgently needed for *data citation*. Other identified requirements have been vocabulary standardisation and the need of open APIs. For example, data and conceptual model standards (e.g. ontologies and vocabularies) strongly reduce the data curation effort and simplify data reuse [29]. The development of minimum information models following the example of MIRIAM [50] would improve data curation. Query interfaces are also in need of a standard [34].

*3.4 Prioritisation and mapping*

The research and innovation topics of Section 3.2 have been mapped to the societal impact they can have in terms of the economic, social and ethical, legal, and political externalities and sectors presented in Section 3.1. The mapping has been done via a review of the BYTE studies and external resources investigating technical requirements, mainly Ref. [2] as this is the main resource from where the BDV SRIA document has evolved. In parallel, the mapping has been done as well at the *BYTE Big data research roadmapping workshop*, where the research topics were also prioritised, extended and revised. Figure 3 evaluates the impact of the research topics to the societal externalities, and Figure 4 shows an independent evaluation of such mapping by the industry and academia experts that participated in the workshop. In the workshop, participants were arranged in groups and codified the priority as high, medium, low or none. Here, we have aggregated the contributions and used a colour scheme from dark blue to white to code the priority of each topic to: top priority (in dark blue) if all or almost all stakeholders (i.e. all working groups but one) agreed the topic to be of high priority; high priority if it was generally considered to be of high priority (most working groups agreed on high priority); medium priority if it was generally considered to be of medium priority (most working groups agreed on medium priority or above); low priority otherwise; and no priority (in white) if all stakeholders agreed the topic has no priority.

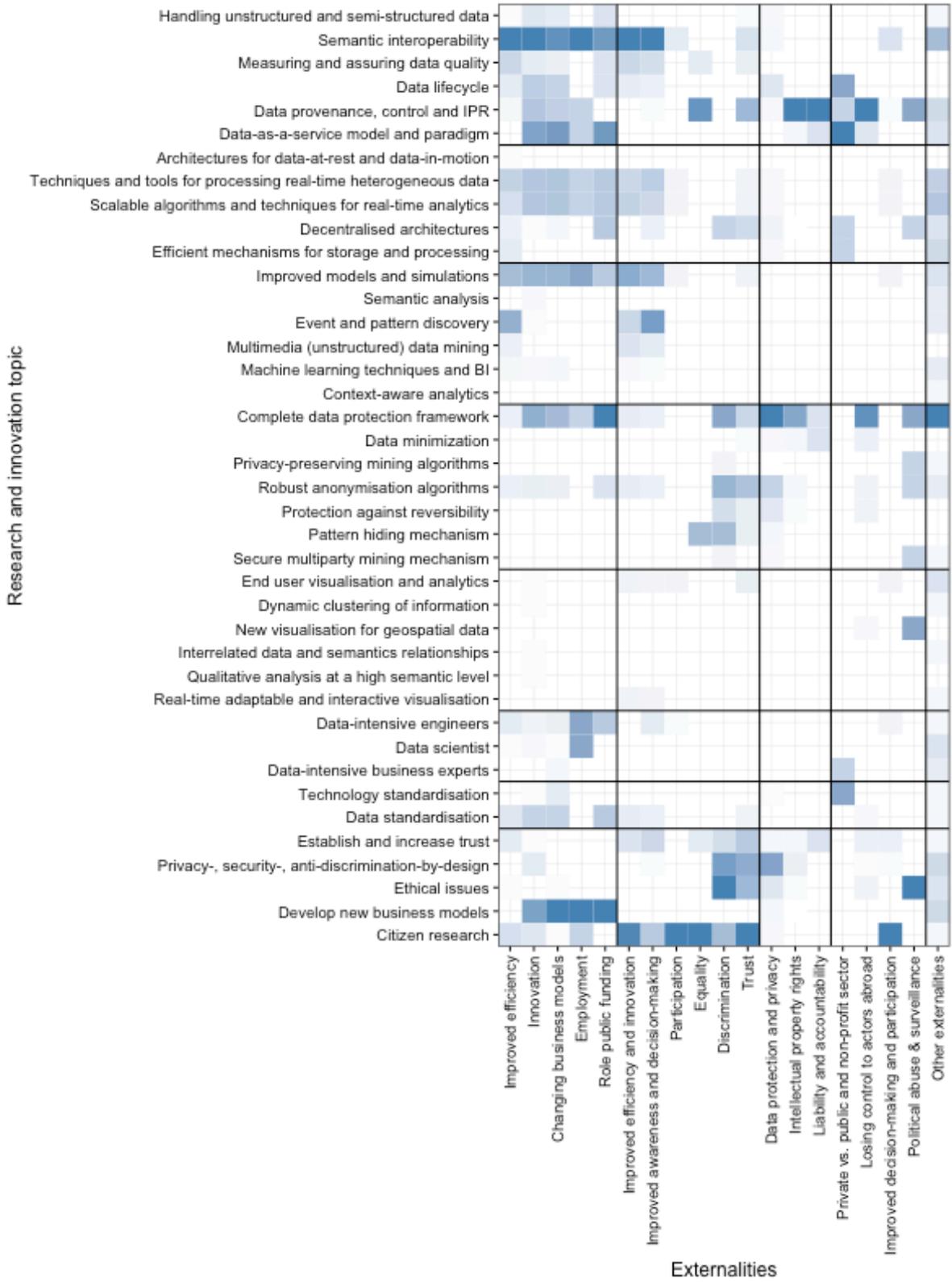

Figure 3. Impact of research in externalities, derived from BYTE analysis and the literature review. Relevance has been normalised by externality: darkest blue corresponds to the most relevant research in the externality. Research topics are grouped in the following areas (top to bottom): data management, data processing, data analysis, data protection, data visualisation, skills development, standardisation, non-technical priorities. Externalities are grouped in four areas (left to right): economic, social and ethical, legal, and political.

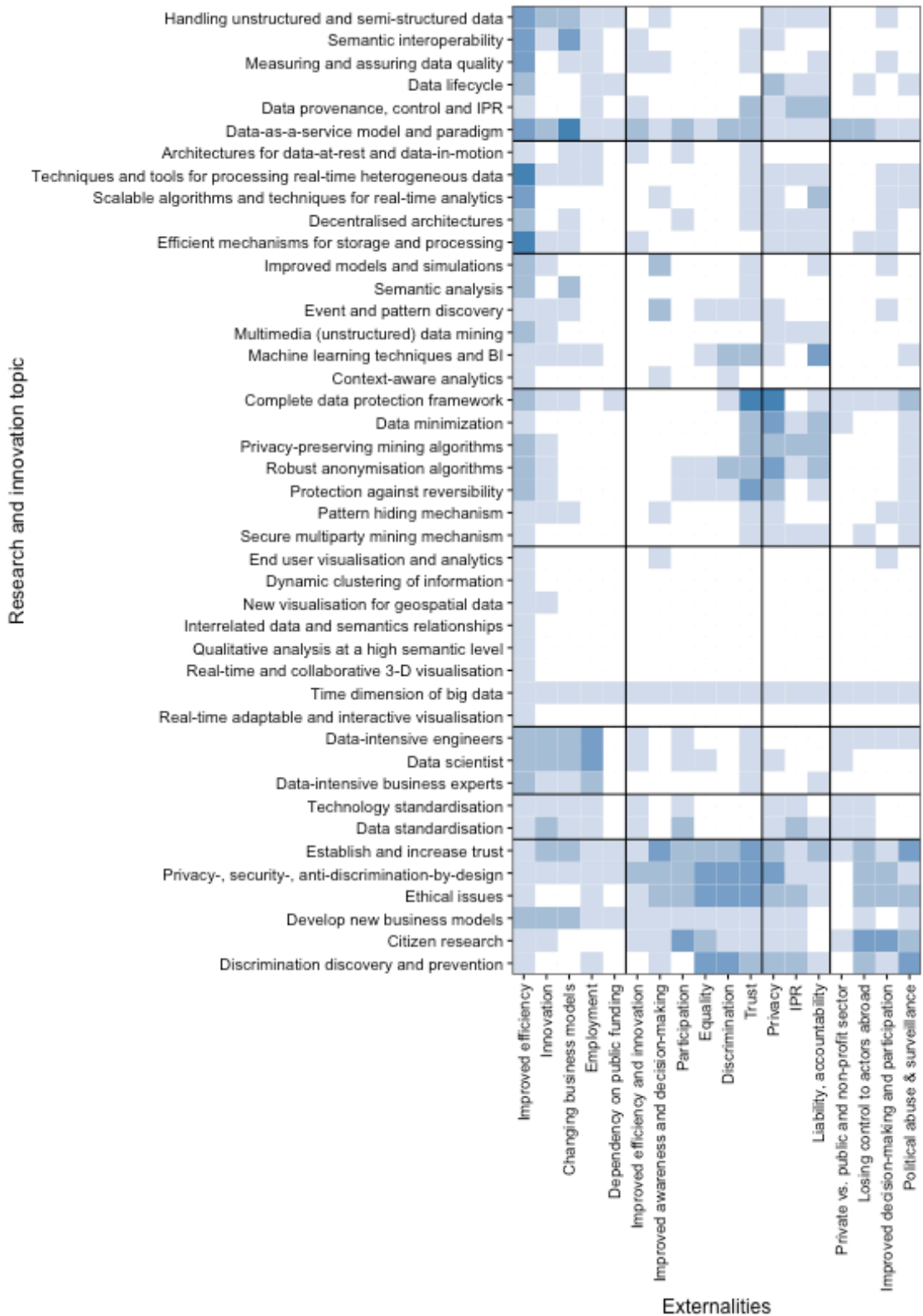

Figure 4. Impact of research in externalities, contributed by stakeholders and community members at the *BYTE Big data research roadmapping workshop*. Relevance has been assessed by workshop participants: top priority (in dark blue) if all or almost all stakeholders agreed the topic to be of high priority; high priority if it was generally considered to be of high priority; medium priority if it was generally considered to be of medium priority; low priority otherwise; and no priority (in white) if all stakeholders agreed the topic has no priority. Research topics are grouped in the following areas (top to bottom): data management, data processing, data analysis, data protection, data visualisation, skills development, standardisation, non-technical priorities. Externalities are grouped in four areas (left to right): economic, social and ethical, legal, and political.

As it can be seen by comparing Figure 3 and Figure 4, the findings of the literature review and the workshop are in good agreement.

To increase the likelihood of technology adoption in the future, the following considerations regarding research topics and their social impact were also put forward by the community in the research roadmapping workshop.

The area of data management is of high priority:

- Without management, there is restricted efficiency and low economic output.
- The data lifecycle is co-dependent on public funding.
- New business models can be created within the **data-as-a-service paradigm**, such as paying for data cleaning.
- **Discrimination** and **trust** are strongly affected by **data management topics**, especially when participation is diminished. Citizens trust is increased by better **data provenance, control and IPR**. On the other hand, businesses trust via innovations in **data-as-a-service model and paradigm**.
- There exists also the risk of losing data to models abroad caused by an inefficient data management.

The area of data processing has moderate to high priority:

- For data-based policy making, it should be considered enforcing at least 3 of the 5 stars of deployment for Open Data (i.e. make data available in a non-proprietary open format). However, it is also worth mentioning that continuing with the current 1-star assessment opens opportunities for other players to create 3-star processing products.
- **Real time efficiency** has moderate to high priority, but legal and especially trust issues require clarification.
- The debate between **decentralized and centralised architectures** needs to be decided. For example, **participation** maybe be positively affected if decentralised, while it was mentioned that the only way for purely **open data** is centralization.

The area of data analytics has also moderate to high priority:

- Research into **multimedia data mining** will lead to **new business models** and innovation, but has important **IPR issues**. Most licenses do not allow for data mining, but the development of **blockchain** may lead to higher participation thanks to an increase of **trust**.
- Smart contracts are a priority. Research in data licensing is also needed.
- Within machine learning techniques for business intelligence, advances in **auditing algorithms** will have a positive impact in **equality**, **discrimination** and **trust** externalities, as well as in **liability and accountability**.
- A common misconception of big data is to ignore "modelling, and instead rely on correlation rather than an understanding of causation" [37] and that with enough data no models are needed [51]. To address this issue, better modelling and simulations, and transparency about the data and the analysis to allow for a validation of the statistical significance of the results are recommended [37]. This includes taking into account design and sample biases.

Data protection:

- In industry, there is still a general fear of sharing data, which is partially compensated by the new value that is added by combining data. Possible solutions that were mentioned by stakeholders are the development of methods, possibly in the design phase and in the line of privacy-by-design, that can increase the trust in the protection of the data, and the development of mechanisms to encourage the emergence of more open business data, such as creating partnerships with public or research organisations that require or encourage open data publication.
- Enhanced cybersecurity captures the positive aspects of trust and privacy externalities.

Data visualisation was viewed as a lower-priority area:

- However, it was brought into attention that **better visualisations and user-friendly interfaces** might decrease the urgent **need of data skills** in the European market.

Non-technical priorities:

- **Data skills** for the general population will capture positive **employment** externalities, especially those connected with data-driven employment offerings and opportunities for economic growth through open data.

- Advances in **data standardisation** support communities and business **partnerships around data**.

**4 Action plan**

In this section we present a timeline of research and innovation topics to tackle the societal externalities, develop the necessary skills and address the standardisation needs of big data in Europe, as well as general recommendations and best practices drawn from the BYTE research and contributions from the community.

*4.1 Timeline*

The topics have been aligned in time in collaboration with stakeholders and the community at the *BYTE Big data research roadmapping workshop*. The timeframe spreads in detail over five years (2017-2021) and includes as well topics to be addressed in the mid- and long-term. This timeline is shown in Figure 5, where research and innovation topics are grouped in the six areas described in the previous section. Figure 5 also visualises the three different means by which these innovations can deliver an impact into the different sectors and externalities: through **standardisation**, **societal impact** and **skills development**, as well as by other means.

We expect research and innovations in these topics to address the negative externalities and deliver positive social benefits in different sectors. The bottommost part of Figure 5 shows the sectors where each topic is expected to have a relevant impact.

In this regard, BYTE has further identified six areas where such positive benefit will be especially relevant. They are:

- data-driven innovations and business models,
- the use of data analytics for large volumes of data to improve event detection, situational awareness, and decision making to e.g. allocate resources efficiently,
- better environmental protection and efficiency and direct social impact to citizens through e.g. individual targeted services,
- the use of big data to enable citizen participation and increase transparency and public trust (this will require efforts to develop data skills among the general public),
- an increased attention paid to privacy and data protection by big data practitioners, and
- big data as a means to identify and combat discrimination.

These benefits are further described in Ref. [3].

*4.2 Best practices*

In order to capture these benefits, several best practices have been suggested by the BYTE project [3,19]. They involve public investments and funding programs to solve the scarcity of European big data infrastructures, promote research and innovation in big data, open more government data and persuade big private actors to release some of their data as well, so data partnerships can be built around them. New data sources and business models also need to be promoted. Interoperability has also been shown to be a key enabling factor. In addition, education policies have to address both the current scarcity of data scientists and engineers, but also the inclusion of data skills in general educational programs.

To address discrimination, equality and trust, privacy-by-design methods should be extended to anti-discrimination-by-design and analogous approaches, and transparency and new accountability frameworks need to be based both on legislation and on a data protection framework. Overall, policy makers, regulators and stakeholders have all an important role in updating legal frameworks, promoting big data practices and developing and incorporating tools into the big data design and practice that address societal concerns.

These best practices can also be followed to capture positive social benefits associated to social externalities. Furthermore, investment in the interoperability of big data is also a key recommended action [19].

Another best practice to address negative social and ethical externalities regarding the risk of discrimination e.g. due to bias in the problem **definition, data mining or training data is to use auditing tools and extend privacy-by-design to anti-discrimination-by-design** [19].

Regarding legal externalities, and besides the need to adapt regulations on a policy level, to address the non-scaling legal frameworks in the context of a high amount of interactions, it is recommended to substitute legal mechanisms based on individual transactions or individual control models with aggregate or collective mechanisms and develop "by-design"-approaches that translate legal objectives into technical requirements [19]. Another recommendation is to develop standardised solutions and a toolbox of legal, organisational and technical means to fine-tune data-flows [19]. Finally, privacy-by-design has to include not only a technical perspective but also legal and organisation safeguards to address the overall capabilities and risks of the systems [19].

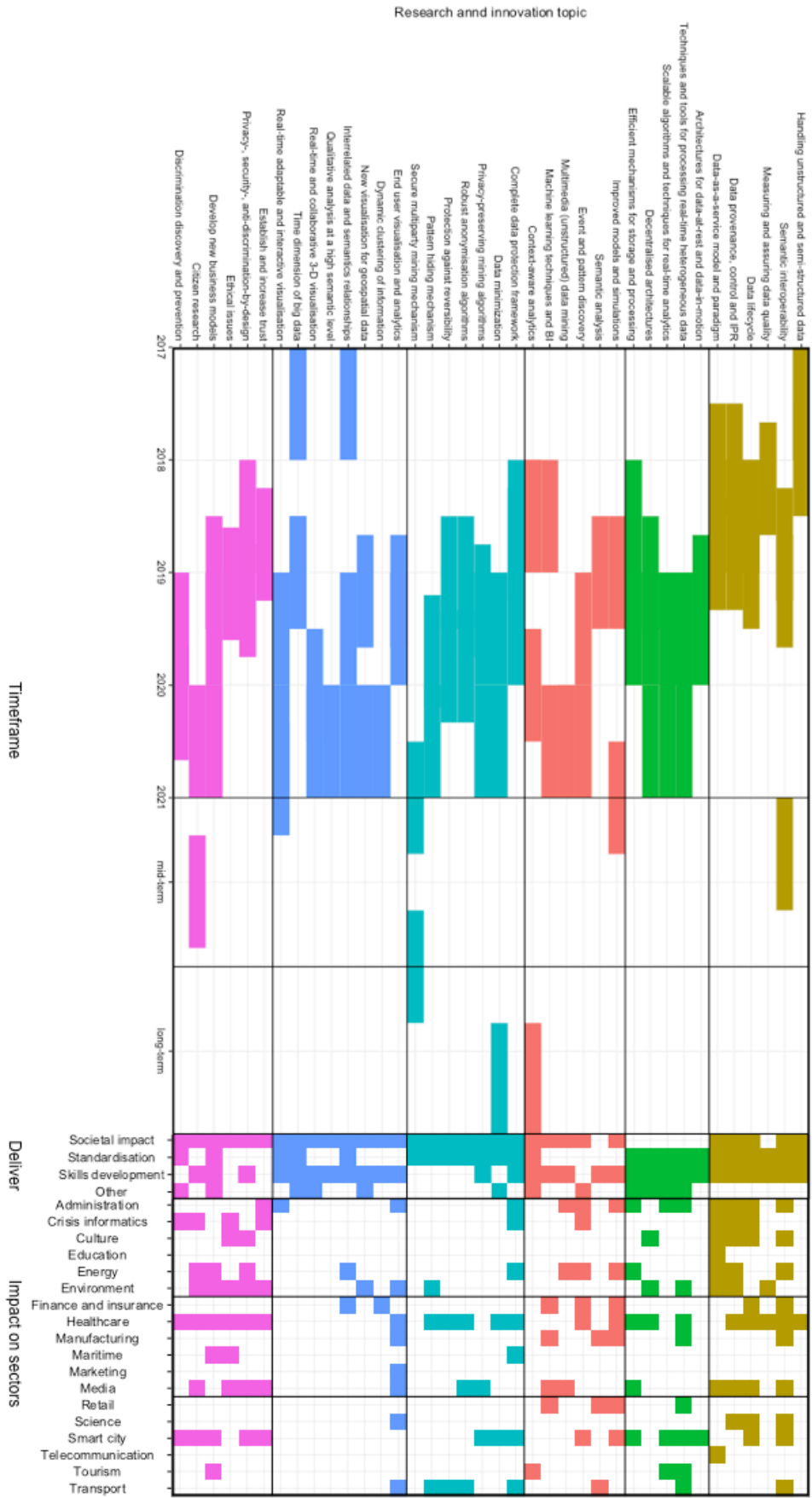

Figure 5. Timeline to address research topics (2017 to 2021, plus mid- and long-term), how they will contribute to deliver societal impact, standardisation and skills development, and their impact to each of the sectors. Research topics are grouped in the following areas (right to left): data management, data processing, data analysis, data protection, data visualisation, non-technical priorities.

## 5 Concluding remarks

This paper has outlined research topics that address societal externalities produced by the use of big data. To deliver social benefits, develop skills and contribute to standardisation, all areas of data management, processing, analysis, protection and visualisation need to be advanced. We have also described how such research efforts are expected to impact on several industry sectors.

Aside from the broad recommendations and timeline to address the research topics presented above, the present roadmap also foresees an annual deeper study of selected sectors to be taken up initially by the BYTE project partners and community members, and by the BYTE community after the project completion. Each year, a group of three sectors will be addressed in detail to produce special recommendations and actions. The goal is that the community is able to present a deeper discussion on what and where are the gaps and challenges each sector faces, and recommend good practices and specific research and policy needs to cover these gaps. For the first year, the environment, healthcare and smart city sectors have been selected by the community as the first to be further studied.

## Acknowledgements

We thank the participants of the *BYTE Big data research roadmapping workshop* and the BYTE consortium members for their insightful contributions. This work was supported by the European Commission through the BYTE project [grant number FP7 GA 619551].